\documentclass[10pt,letterpaper]{article}
\usepackage[top=0.85in,left=2.75in,footskip=0.75in]{geometry}

\usepackage{amsmath,amssymb}

\usepackage{changepage}

\usepackage{textcomp,marvosym}

\usepackage[numbers]{natbib}

\usepackage{nameref,hyperref}

\usepackage[right]{lineno}

\usepackage[nopatch=eqnum]{microtype}
\DisableLigatures[f]{encoding = *, family = * }

\usepackage[table]{xcolor}

\usepackage{array}

\newcolumntype{+}{!{\vrule width 2pt}}

\newlength\savedwidth

\raggedright
\usepackage[aboveskip=1pt,labelfont=bf,labelsep=period,justification=raggedright,singlelinecheck=off]{caption}

\makeatletter
\renewcommand{\@biblabel}[1]{\quad#1.}
\makeatother

\usepackage{lastpage,fancyhdr,graphicx}
\usepackage{epstopdf}
\fancyheadoffset[L]{2.25in}
\fancyfootoffset[L]{2.25in}
\begin{document}
\vspace*{0.2in}

\begin{flushleft}
{\Large
\textbf{Machine Learning for Climate Downscaling in CORDEX: Opportunities and Challenges} 
}
\newline
Neelesh Rampal\textsuperscript{1},
Jos\'e M. Guti\'errez\textsuperscript{2},
Jorge Ba\~no-Medina\textsuperscript{2},
Maria Laura Bettolli\textsuperscript{3},
Antoine Doury\textsuperscript{4},
Fatima Driouech\textsuperscript{5},
Ram\'on Fuentes-Franco\textsuperscript{6,7,8},
Peter B. Gibson\textsuperscript{1},
Andrew Orr\textsuperscript{9},
Pedro M. M. Soares\textsuperscript{10},
Peter A. G. Watson\textsuperscript{11}
\\
\bigskip
\textbf{1} Earth Sciences New Zealand (Formerly NIWA), New Zealand
\\
\textbf{2} Instituto de F\'isica de Cantabria, CSIC--Univ.\ de Cantabria, Santander, Spain
\\
\textbf{3} Universidad de Buenos Aires--CONICET--IFAECI, Argentina
\\
\textbf{4} Centre National de Recherches M\'et\'eorologiques (CNRM), Universit\'e de Toulouse, M\'et\'eo-France, CNRS, Toulouse, France
\\
\textbf{5} University Mohammed VI Polytechnic, Morocco
\\
\textbf{6} Rossby Centre, SMHI, Norrk\"oping, Sweden
\\
\textbf{7} Bolin Centre for Climate Research, Stockholm University, Stockholm, Sweden
\\
\textbf{8} Swedish Centre for Impacts of Climate Extremes, Uppsala University, Uppsala 75236, Sweden
\\
\textbf{9} British Antarctic Survey, Cambridge, UK
\\
\textbf{10} IDL -- Instituto Dom Luiz, Faculdade de Ci\^encias, Universidade de Lisboa, 1749-016 Lisbon, Portugal
\\
\textbf{11} School of Geographical Sciences, University of Bristol, UK
\\
\bigskip

* E-mail: neelesh.rampal@earthsciences.nz

\end{flushleft}


%
\section*{Abstract}
The Coordinated Regional Climate Downscaling Experiment (CORDEX) has been instrumental in advancing regional climate modeling and in enabling the provision of high-resolution climate information for climate services worldwide. However, the high computational cost of regional climate models (RCMs) remains a significant barrier to producing large ensembles of continental-scale projections at fine spatial resolution. The computational barrier also hinders the ability of many regions (particularly across the Global South) to participate in these efforts. Over the past two decades, machine learning (ML) has emerged as a promising and computationally efficient alternative for climate downscaling, yet significant challenges remain in systematically assessing the performance, robustness and applicability of the diverse range of ML-based approaches across climatic regions. The current lack of coordination further hampers progress, and ultimately trust in the projection output, underscoring the need for improved standardization, benchmarking, and shared best practices. The CORDEX Task Force on Machine Learning was established to address these challenges by coordinating ML-based downscaling activities and ensuring alignment with other CORDEX activities. A key activity was CORDEX-ML-Bench, a lasting set of experiments, protocols, and datasets that supported an intercomparison of 40 models from over 20 institutions and yielded important findings on model comparison, extrapolation, and transferability. This paper presents the Task Force's outcomes: proposed terminology for a field with heterogeneous naming and framing, a mapping of the current landscape, and advances in benchmarking and evaluation. It also identifies open scientific challenges and future directions for a follow-up Task Team within CORDEX.


%

\clearpage
\newgeometry{top=0.85in,left=1in,right=1in,footskip=0.75in}

\section{Introduction}
\label{sec:introduction}

By providing detailed regional climate information, the Coordinated Regional Climate Downscaling Experiment (CORDEX) has played a central role in advancing scientific understanding of climate variability and change at regional scales over all areas of the planet \citep{gutowski2016,jacob2020,diezsierra2022,orr2023}. Its outputs are widely used in international synthesis assessments such as the IPCC Sixth Assessment Report Atlas \citep{atlas2021}, as well as in climate impact assessments, adaptation planning, and climate services \citep{diezsierra2022}. These efforts directly support risk analysis and decision-making across multiple sectors on regional to local scales. A key strength of CORDEX lies in its coordinated use of high-resolution Regional Climate Models (RCMs) for dynamical downscaling of low-resolution Global Climate Models (GCMs). This has fostered global communities of practice \citep{gilbert2026} while providing common protocols that enable robust intercomparison studies and enhance the usability of regional climate information \citep{bukovsky_2026}.

Despite its success, CORDEX faces several substantial challenges, many of which stem from the computational demands and practical limitations of running RCMs, especially at kilometer-scales. First, the high computational cost restricts the production of large ensembles, limiting the sampling of uncertainty across GCM--RCM combinations, emission scenarios, and model realisations, thereby preventing a full representation of plausible climate futures and associated uncertainty \citep{hawkins2009,evin2021,lehner2023}. Second, generating coordinated ensembles with multiple RCMs remains particularly difficult in resource-limited regions, resulting in inequities across CORDEX domains. Subsequently, some of the regions most vulnerable to climate change remain underrepresented in coordinated downscaling efforts. Third, the transition toward kilometre-scale convection-permitting modeling, while promising, remains computationally prohibitive for generating sufficiently century-long large ensemble simulations.

Empirical--statistical downscaling (ESD), also known as observational downscaling, is a long-established approach that trains a statistical model to map coarse-resolution reanalysis fields to local-scale observations, and then applies the trained model to GCM output \citep{wilby1998}. As a computationally efficient alternative to running GCM--RCM combinations, ESD can alleviate many of these constraints, enabling the production of large ensembles and broader regional participation at a fraction of the cost. CORDEX has recognized this potential and incorporated ESD into coordinated evaluation and intercomparison exercises at continental scale \citep{maraun2015,gutierrez2019}. In practice, however, ESD applications have largely remained national or subnational, and concerns persist about their extrapolation beyond training conditions and their skill on extremes \citep{gutierrez_reassessing_2013, rampal2024enhancing}. Because ESD applications have been localized and often case-specific, comprehensive intercomparisons across domains remain limited, as does the coordinated production of statistically downscaled climate information. As a result, most available products are still global or regional projections adjusted with simple bias-adjustment methods \citep[e.g.,][]{yakubu2025}.

Recent advances in the field of machine learning (ML) offer a promising avenue to address several of these challenges \citep{bano2020, doury2023, rampal2024enhancing, GIBSON2026}. State-of-the-art ML approaches have been widely applied as modern extensions of ESD, combining flexible non-linear architectures and large datasets with the core principle of learning statistical relationships between large-scale predictors and local climate variables \citep{rampal2024enhancing, beucler2026artificial}. More recently, a growing body of work has explored RCM downscaling emulators, in which ML-based downscaling models are trained on RCM simulations rather than observations, allowing longer training periods that span both historical and future climates \citep{doury2023, kendon2025}. Once trained, these ML-based models run orders of magnitude faster than RCMs, making it feasible to generate detailed continental-scale climate information and large ensembles at low cost, enabling more comprehensive uncertainty quantification across internal variability, model uncertainty, and emission scenarios \citep{rampal2024enhancing}. Beyond efficiency, ML methods offer advantages in scalability and accessibility and could lower entry barriers for research groups in resource-limited regions.  In this sense, ML should be viewed not as a standalone replacement for dynamical downscaling---since physics-based simulations remain essential for generating physically consistent training data and exploring novel climate regimes---but as a complementary tool capable of extending the reach, inclusivity, and efficiency of coordinated regional climate information production. Nevertheless, ML approaches also face important and unique limitations, including challenges related to generalization under climate change, physical consistency, and dependence on training data \citep{rampal2024enhancing, kendon2025, rampal2026cordex, GIBSON2026, Reyes-Elgueta_2026}.

Recognizing both these opportunities and challenges, CORDEX established a Task Force on Machine Learning (TF-ML) which has subsequently evolved into a Task Team. The overarching goal is to coordinate ML-based downscaling activities and foster climate downscaling collaboration between ML researchers and climate scientists across a diverse range of methods, spanning classical statistical learners (e.g.  random forests and multiple linear regression) through to state-of-the-art deep neural networks, generative models, and physics-informed architectures. A central objective is the establishment of standards, benchmarks, and best practices to ensure robust evaluation and responsible application of ML-based downscaling methods. To maximize the potential benefits, these activities need to be done in close coordination with CORDEX regional activities, Flagship Pilot Studies (FPS), and other Task Teams, particularly those focused on convection-permitting modeling, as well as relevant external initiatives. Until recently, climate downscaling lacked a standardised benchmark framework comparable to WeatherBench for weather prediction \citep{rasp2020, rasp2024weatherbench} and ClimateBench for AI-based global climate emulators \citep{watson2022}. The CORDEX Task Force on Machine Learning (TF-ML) has since addressed this gap by developing CORDEX-ML-Bench \citep{rampal2026cordex}, which offers a common basis for evaluating and comparing downscaling methods and lays the groundwork for coordinated progress.

This paper presents the outcomes of the TF-ML, including a mapping of the current landscape of ML-based downscaling methods and advances in benchmarking and evaluation frameworks. Building on these results, we assess the key methodological and scientific challenges facing ML-based downscaling and identify critical gaps that must be addressed for their successful integration within the CORDEX framework. Finally, we propose future directions and actions to be pursued through a dedicated Task Team follow-up, aimed at sustaining and expanding coordinated ML-based climate downscaling activities across CORDEX domains.

\section{The Landscape of ML-Based Climate Downscaling}
\label{sec:landscape}

\begin{figure}[tbp]
\centering
\includegraphics[width=\linewidth]{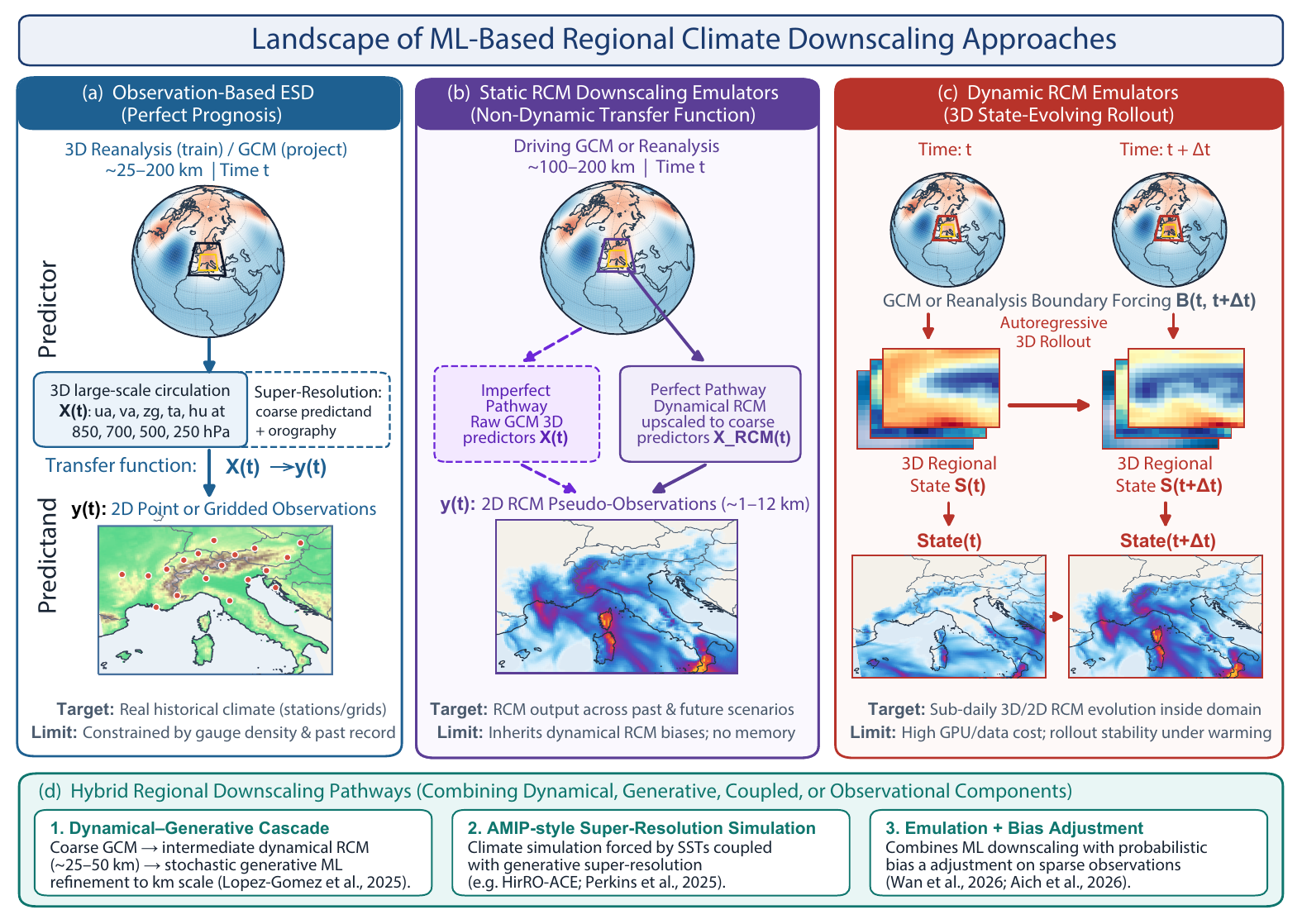}

\caption{Schematic representation of ML-based regional climate downscaling approaches, highlighting the distinction between \textbf{(a)} observation-based ESD (blue), \textbf{(b)} static RCM downscaling emulators spanning imperfect (raw GCM predictors) and perfect (upscaled RCM predictors) pathways (violet), \textbf{(c)} dynamic (3D autoregressive) RCM emulators driven by GCM or reanalysis boundary forcing (crimson), and \textbf{(d)} hybrid regional downscaling pathways. While ESD and static RCM emulators estimate high-resolution surface predictands directly from coarse-resolution predictors at the same time $t$, dynamic RCM emulators evolve the 3D regional state at time $t + \Delta t$ from the regional state at time $t$. Regional maps show actual CORDEX-ML-Bench European Alps topography and daily precipitation.}
\label{fig:landscape}
\end{figure}

\subsection{ML-Based Downscaling Approaches}
\label{sec:approaches}

ML-based regional climate downscaling follows two main paradigms: observation-based empirical statistical downscaling (ESD) and regional climate model (RCM) emulators. Both typically use similar ML architectures but differ in their training data: ESD is trained on observations, whereas emulators are trained on RCM simulations. This difference matters for their respective strengths and limitations \citep{rampal2024enhancing, kendon2025}. Observation-based ESD models are trained on observational datasets (station-based or gridded) using large-scale reanalysis fields as predictors, an approach often called perfect prognosis (see Figure \ref{fig:landscape}a). Because they mimic observations on the ground, they can capture historical variability with reasonable accuracy. A simpler form of ESD is statistical super-resolution (SR), in which high-resolution observational fields are reconstructed from their spatially coarsened counterparts, without explicitly using large-scale atmospheric predictors. This approach focuses primarily on recovering fine-scale spatial patterns and variability from coarse-resolution fields. Closely related are bias adjustment methods \citep{potter2023, maraun2016bias, maraun2013bias}, which also adjust climate model output toward observations but, unlike ESD trained on paired samples, aim to match distributions rather than individual realizations \citep{aich2026conditional, hess2023deep}.

Observation-based ESD models have several important limitations. First, they are fundamentally constrained by the spatial coverage (e.g., rain gauges), quality, and length of the observational record \citep{Reyes-Elgueta_2026}. This raises broad concerns about their ability to extrapolate reliably to climate conditions outside their training period \citep{hernanz2022, gonzalez2025, wang-mitigating_2026}. In contrast, RCM emulators are trained to replicate the downscaling relationship of a specific RCM by learning to map low-resolution atmospheric predictors to high-resolution RCM surface variables, which act as synthetic ``pseudo-observations'' (see Figure \ref{fig:landscape}b). The predictors can be derived either from coarsened RCM fields (``perfect'' training framework) or directly from raw GCM outputs (``imperfect'' training framework) \citep[e.g.,][]{doury2023,bano2024,kendon2025}. Because RCM emulators can be trained on simulations spanning future emission scenarios, they are better equipped to extrapolate to warmer climates \citep{rampal2024extrapolation}, though they inevitably inherit physical biases present in the underlying dynamical models unless paired with subsequent observational post-processing or hybrid bias adjustment \citep{girona2025,antonio2025}.

More recently, dynamic (autoregressive) approaches have been developed for RCM emulators (see Figure \ref{fig:landscape}d), analogous to those adopted in data-driven weather forecasting and global climate emulators \citep{lang2024,vankatwyk2026}. Rather than mapping coarse predictors to surface variables independently at each step, these models learn to predict the 3D regional state of the system at time $t + \Delta t$ as a function of the regional state at time $t$, such that they are more closely aligned with emulating the prognostic temporal integration performed by numerical limited-area models \citep{sha2026}. An RCM is run nested within a driving physics-based GCM or reanalysis, which supplies its time-varying lateral boundary and radiative forcings, and resolves finer-scale regional processes. A dynamic RCM emulator is more similar to this setup and predicts the next 3D regional state from the current regional state and the same driving forcings (see Section~\ref{sec:memory} for temporal memory). To clarify terminology across these designs, we distinguish between \emph{static} (non-dynamic) RCM downscaling emulators, \emph{dynamic} (3D state-evolving autoregressive) RCM emulators, and \emph{hybrid} RCM emulators---which combine dynamical and ML components, such as cascading an intermediate-resolution RCM with a generative downscaler \citep{boe2023,lopezgomez2025}, embedding physical conservation layers \citep{harder2022}, or coupling emulation with statistical bias adjustment \citep{girona2025,antonio2025}---retaining \emph{RCM emulator} as the overarching term.

Figure~\ref{fig:landscape}a-d presents a schematic of the ML-based downscaling landscape described above. Both observation-based ESD (blue; \textbf{a}) and static RCM downscaling emulators (violet; \textbf{b}) operate as cross-scale transfer functions that map 3D coarse-resolution atmospheric predictors from reanalysis or a driving GCM (or coarsened RCM fields in the perfect pathway), together with fine-scale static covariates, directly to 2D high-resolution surface variables at the same time step without integrating a 3D regional state. In contrast, dynamical RCM emulators (crimson; \textbf{c}) predict the 3D regional atmospheric state at time $t + \Delta t$ from the state at time $t$, conditioned on lateral boundary forcing from a driving GCM or reanalysis \citep[e.g.,][]{sha2026}, and advance in time through autoregressive rollout. Hybrid pathways (bottom bar) combine elements of all paradigms. These include dynamical generative approaches \citep{lopezgomez2025}, AMIP-style rollouts of neural weather models forced by sea surface temperatures (SSTs) \citep{henn2026aimip}, and super-resolution with bias adjustment \citep{aich2026conditional, wan2026regional}, the latter being similar to observation-based ESD. 

The CORDEX ML benchmark provides guidance on data preprocessing, normalisation, model training and evaluation \citep{rampal2026cordex}. Training data is typically normalised with $z$-scores for predictors and temperature, and log, square-root or quantile transforms for precipitation, which helps balance the loss across variables \citep{doury2023,rampal2026cordex}. Trained models should then be compared against simple baselines using a common set of metrics that go beyond pointwise errors (Section~\ref{sec:benchmark}), before being applied to unseen GCMs using the training grid and normalisation statistics.

A related open question concerns the minimum training data volume and climate-state diversity needed for reliable extrapolation, particularly for costly convection-permitting simulations. Recent work shows that the required training length varies across models and metrics, with longer records needed to
capture precipitation extremes \citep{gibson2026longer}. Coordinated benchmarks such as CORDEX-ML-Bench are well positioned to address these questions systematically.

\subsection{A Timeline of Recent Advances in ML-Based Downscaling}
\label{sec:timeline}

\begin{figure}[tbp]
\centering
\includegraphics[width=1.03\linewidth]{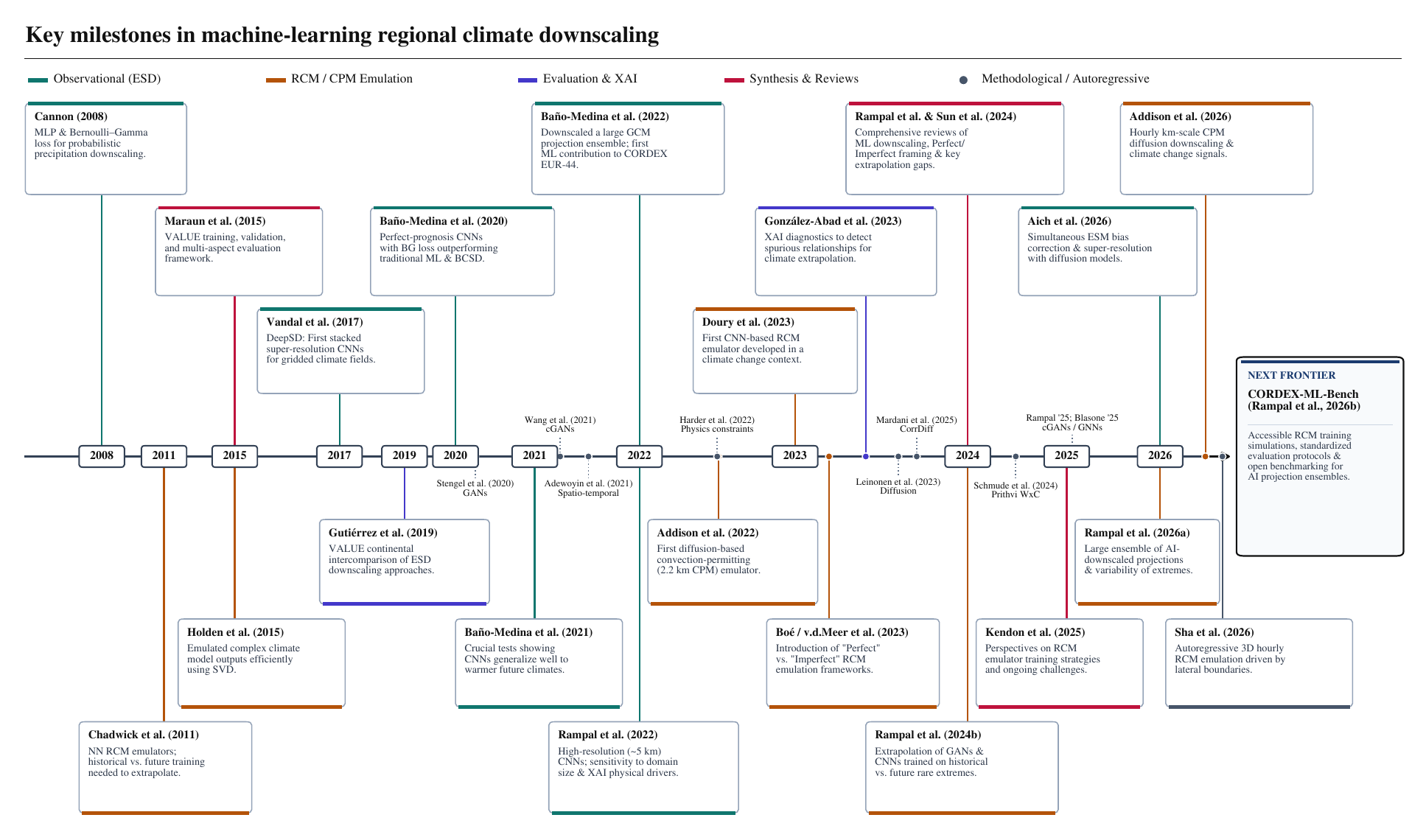}

\caption{A timeline of key publications in ML-based regional climate downscaling.}
\label{fig:milestones}
\end{figure}

Having outlined the main approaches, we now trace how the field has evolved. Figure~\ref{fig:milestones} presents a timeline of key publications in ML-based downscaling, spanning both observation-based ESD and RCM emulators, and their main contributions.


Both ESD and RCM emulation initially drew heavily from conventional ML methods, with relatively simple and shallow artificial neural networks (ANNs) included in statistical downscaling intercomparisons as early as the late 1990s \citep{wilby1998}, alongside singular value decomposition, support vector machines, random forests, and Gaussian processes \citep{cannon2008,chadwick2011,holden2015,girona2025}. A number of insights from these early applications have remained informative. For example, \citet{cannon2008} introduced the use of a Bernoulli--Gamma (BG) log-likelihood for precipitation, a stochastic formulation that proved important in subsequent deep learning approaches for predicting precipitation and extremes. \citet{chadwick2011} highlighted the importance of training on future climates to improve extrapolation, a challenge that has remained central to the field and continues to inform the development of more sophisticated approaches, including with RCM emulators. 

\begin{figure}[tbp]
\centering
\includegraphics[width=\linewidth]{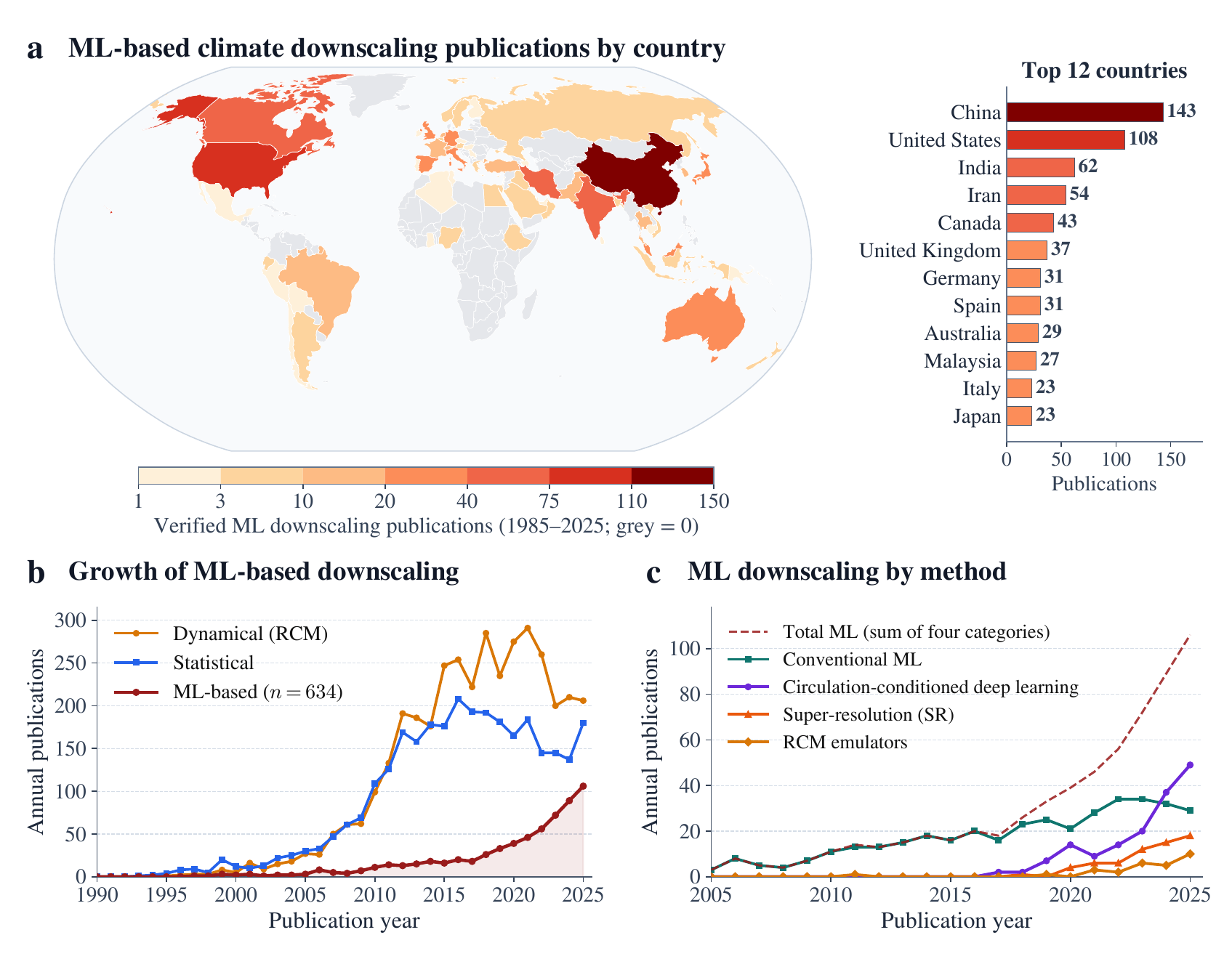}
\caption{\textbf{(a)} Geographical distribution of ML-based downscaling publications by author institutional affiliation, across Earth, environmental, computational, engineering, energy, and mathematical science subject areas. \textbf{(b)} Annual publication counts for dynamical (RCM), conventional statistical, and ML-based downscaling. \textbf{(c)} Temporal evolution of ML downscaling sub-categories: deep learning (CNNs, GANs, diffusion models, Transformers), super-resolution (SR), RCM emulators, and conventional ML (random forests, support vector machines, Gaussian processes, and shallow ANNs). Counts may be conservative, as searches were restricted to titles, abstracts, and keywords.}
\label{fig:bibliometrics}
\end{figure}

The success of convolutional neural networks (CNNs) in computer vision applications \citep{krizhevsky2012}, the subsequent adoption of U-Net architectures \citep{ronneberger2015}, and more broadly the rise of deep learning around the year 2015 have seen a new wave of ML-based downscaling studies, as also illustrated in Figure~\ref{fig:bibliometrics}b-c. These early deep learning approaches drew heavily on super-resolution (SR) techniques from the computer vision community \citep{vandal2017}, training models to predict high-resolution variables solely and directly from coarsened versions of those same surface variables, such as coarsened fields from atmospheric reanalyses or RCM simulations. Although extensively explored, a limitation of these techniques is that they generally do not incorporate large-scale atmospheric 3D predictors and therefore do not condition the downscaling function on large-scale circulation. Concerns also remain regarding the transferability of SR approaches to free-running GCM simulations, as surface variables such as precipitation tend to be less realistically represented than large-scale free-tropospheric variables when compared with reanalysis \citep{rampal2024enhancing}. These limitations have contributed to the relatively limited adoption of SR approaches within the climate downscaling community \citep{rampal2024enhancing}.



Early approaches with shallow ANNs could not directly ingest high-dimensional spatial grids and therefore relied on human-guided feature engineering such as principal components or local grid-point selection \citep{hobeichi2023using}, limiting their ability to capture spatial relationships---an important shortcoming for variables such as precipitation, where synoptic spatial gradients govern local intensity \citep{rampal2022high}.

The adoption of CNNs addressed a limitation of earlier models, namely their reliance on manual feature engineering, such as principal component analysis or local grid-point selection \citep{hobeichi2023using}. By  automatically learning spatial dependencies from gridded 3D atmospheric fields, CNNs reduced manual intervention in predictor design \citep{bano2020}, and many studies using CNN-based approaches have demonstrated clear improvements over conventional methods in several continental- and regional-scale applications \citep[e.g.,][]{bano2022,balmaceda2023,soares2024,johannsen2024, rampal2022high}. However, CNNs trained with deterministic regression-based loss functions such as mean squared error (MSE) tend to smooth fine-scale details and underestimate extremes \citep{rampal2025reliable}. This limitation has motivated a shift toward stochastic generative approaches, including Generative Adversarial Networks (GANs), diffusion models and flow matching, which can predict multiple plausible high-resolution realizations from coarse-resolution predictors and show promise in better capturing extremes and fine-scale variability, including at convection-permitting scales \citep[e.g.,][]{stengel2020,addison2022,addison2026}. While many GANs and diffusion models build on CNN backbones, Transformer architectures \citep{vaswani2017,schmude2024,bodnar2025} are gaining traction, with attention mechanisms better capturing dependencies across multiple spatial scales \citep[e.g.,][]{bailie2024quantile}, a known limitation of standard convolutions. Fourier Neural Operators represent another promising direction with similar potential advantages \citep{pathak2022}.

While ML-based ESD approaches have shown promising performance \citep[see][]{gonzalez2025,bettolli2025}, their strong reliance on past observations raises concerns about their ability to extrapolate to future climates. This limitation has driven growing interest in RCM emulation, which aligns directly with CORDEX's core mission by enabling the community to populate missing combinations in sparse GCM--RCM--scenario matrices and expand ensemble sizes, and the recent CORDEX benchmark has shown that emulators can reproduce key climate statistics and even certain extremes \citep{rampal2026cordex}. 

Leading technology laboratories and public--private partnerships have also recently begun developing climate downscaling approaches, including foundation models (large neural networks pretrained on broad datasets and adapted to many downstream tasks) as well as diffusion-based frameworks.
Examples include diffusion and flow-matching convection-permitting downscalers within NVIDIA's Earth-2 platform such as CorrDiff \citep{mardani2025,fotiadis2024}, Earth-system foundation models with downscaling capabilities such as NASA--IBM's Prithvi~WxC \citep{schmude2024} and Microsoft's Aurora \citep{bodnar2025}, cloud-hosted statistically downscaled projection initiatives \citep{gergel2024}, and hybrid dynamical--diffusion regional downscaling frameworks \citep{lopezgomez2025}. Other approaches, such as the Allen Institute for AI’s Hiro-ACEv2 \citep{perkins2025hiro, kwa2026extreme} and NVIDIA’s Climate in a Bottle \citep{brenowitz2025climate}, differ from conventional ML downscaling: they simulate weather directly, as in AI-based forecasting, while also producing fine-scale fields through super-resolution. As with ESD, because these models are typically trained on reanalysis data, they may struggle to respond realistically to climate change signals. Training them on future climate simulations may improve their responses, though this remains underexplored. 

To further characterize the evolution and global footprint of ML-based downscaling approaches within the broader regional climate literature, we conducted a bibliometric analysis of peer-reviewed articles and reviews (1985--2025) indexed across Earth and planetary sciences, environmental science, computer science, engineering, energy, and mathematics (Figure~\ref{fig:bibliometrics}). Publications were identified by searching titles, abstracts, and keywords for downscaling terms combined with core climate variables (e.g., temperature, precipitation, wind, or extremes), excluding weather-forecasting studies. They were then classified as dynamical RCM downscaling, conventional statistical downscaling, or ML-based downscaling, with the latter subdivided into deep learning transfer functions, super-resolution, RCM emulation, and conventional ML methods (e.g., random forests, support vector machines, Gaussian processes, and shallow ANNs). ML-based downscaling research is concentrated in China, the United States, Western Europe, India, Canada, and Australia, with far fewer contributions from parts of Africa, Central and South America, and Southeast Asia (Figure~\ref{fig:bibliometrics}a). Publications have grown rapidly since 2017 with the adoption of deep learning, approaching the annual volume of conventional statistical (including bias adjustment) and dynamical RCM downscaling by 2024--2025, both of which appear to have plateaued (Figure~\ref{fig:bibliometrics}b). This growth reflects a shift from conventional ML and pure image super-resolution toward circulation-conditioned deep learning, generative models, and RCM emulators (Figure~\ref{fig:bibliometrics}c).

\section{Benchmarking for Standardised Evaluation}
\label{sec:benchmark}

Despite recent advances in ML-based climate downscaling, progress has been difficult to quantify because existing studies use different datasets, predictors, temporal resolutions, and geographic domains, which precludes meaningful intercomparison. CORDEX has extensive experience coordinating downscalinf efforts, notably through the ESD intercomparison experiment supported by the VALUE evaluation framework \citep{gutierrez2019}, which demonstrated the benefits of coordinated and transparent protocols. However, few comparable efforts have followed, and the emergence of ML-based approaches calls for new experiments and metrics beyond those designed for the conventional ESD methods assessed in VALUE.

Standardized evaluation is a prerequisite for integrating ML-based approaches into climate services, as it underpins scientific credibility, physical consistency and, ultimately, ethical accountability. Benchmarks should therefore define dedicated experiments and metrics to assess fitness for purpose, including skill for extremes and, critically, extrapolation to unseen climates, where ML models often perform poorly \citep{hernanz2022,gonzalez2025}. They should also test transferability from ``perfect'' predictors (reanalysis or coarsened RCM fields) to imperfect GCM predictors with biases and internal states that differ from those of perfect predictors; skill in the perfect setting alone therefore does not establish fitness for purpose.

In the ML research community, intercomparison is typically structured around benchmarks designed to provide standardized datasets, metrics, and evaluation protocols for rapid, reproducible, and fair comparison \citep{dueben2022,ullrich2025}, accelerating methodological progress and establishing good practices. Benchmarks exist for meteorological downscaling \citep[e.g.,][]{langguth2024, harder2025rainshift}, for neural weather models \citep[e.g.,][]{rasp2024weatherbench}, and for AI-based global climate model emulators \citep{watson2022, watson2026climatebench, henn2026aimip}. To date, however, none provides a comprehensive framework for downscaling climate change applications, particularly for assessing extrapolation to future climates and transferability across driving models.

\begin{figure}[tbp]
\centering
\includegraphics[width=0.75\linewidth]{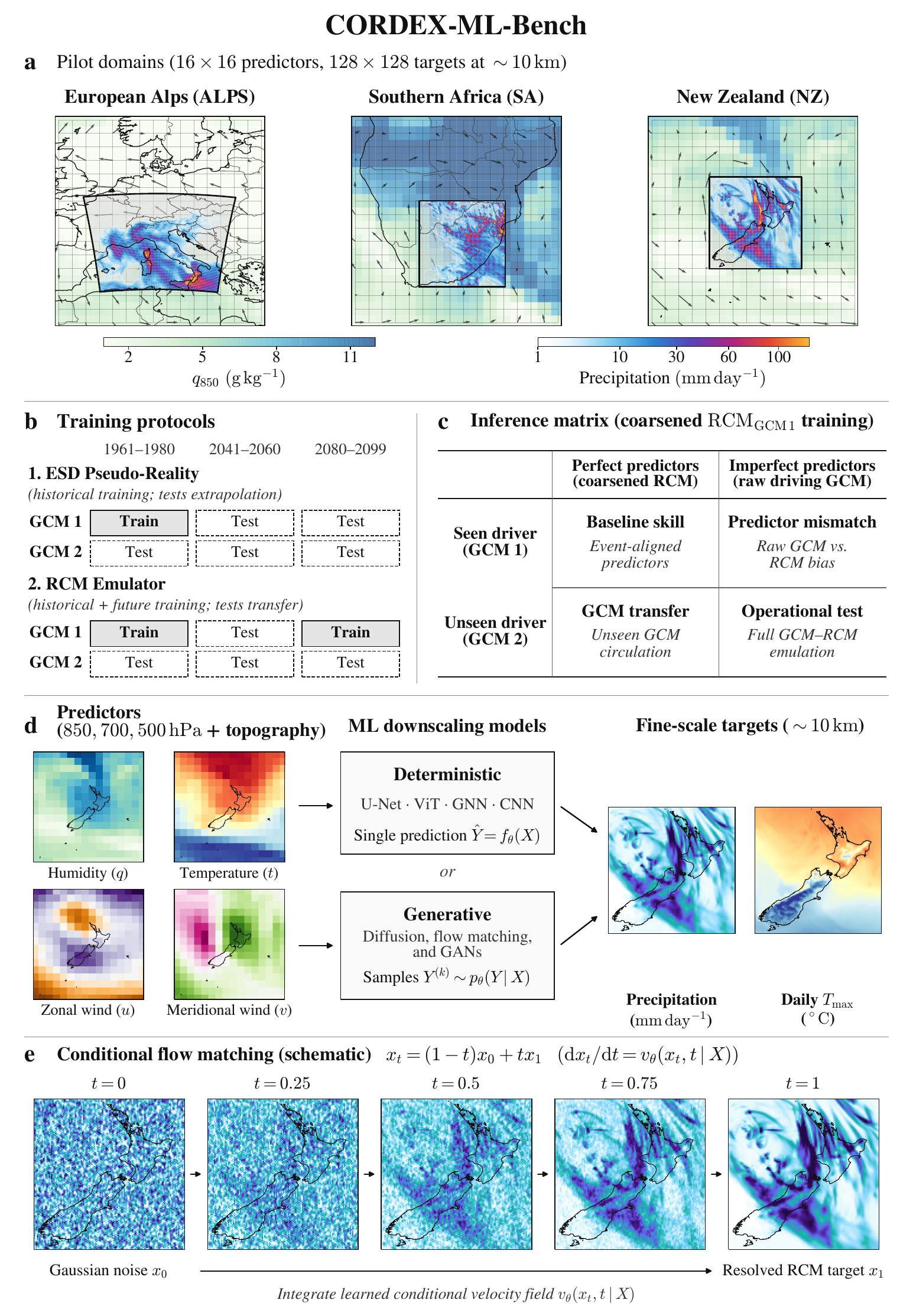}
\caption{Overview of the CORDEX-ML-Bench framework \citep{benchmarkdata2025,benchmark2026}. \textbf{(a)} Daily predictor--predictand snapshots across the three pilot domains (European Alps, Southern Africa, and New Zealand), showing coarse $16\times 16$ 850\,hPa specific humidity ($q_{850}$) and wind vectors surrounding the nested $128\times 128$ ($\sim 10\,\text{km}$) RCM daily precipitation target. \textbf{(b)} Standardized training protocols (historical-only ESD pseudo-reality vs.\ historical + end-of-century RCM emulation), with training on $\text{GCM~1}$ and testing across periods and driving GCMs. \textbf{(c)} Two-by-two inference matrix, for models trained on coarsened RCM output driven by $\text{GCM~1}$, crossing predictor provenance (Perfect coarsened RCM vs.\ Imperfect raw driving GCM) with driving-GCM identity (Seen $\text{GCM~1}$ vs.\ Unseen $\text{GCM~2}$ for the same target RCM). \textbf{(d)} Example of the downscaling task for New Zealand: coarse predictors (humidity, temperature, and zonal and meridional wind at 850, 700, and 500\,hPa, plus static topography) are mapped by deterministic (e.g., U-Net, ViT, GNN, CNN) or generative (e.g., diffusion, flow matching, GAN) models to fine-scale ($\sim 10\,\text{km}$) precipitation and daily maximum temperature. \textbf{(e)} Illustration of a generative flow/diffusion trajectory, in which a learned conditional velocity field transforms Gaussian noise ($t=0$) into the resolved RCM precipitation target ($t=1$).}
\label{fig:benchmark}
\end{figure}

Recognizing these gaps, the CORDEX ML Task Force (TF-ML) conducted a first coordinated benchmarking activity, CORDEX-ML-Bench \citep{rampal2026cordex}, designed explicitly for climate applications. It aims to provide a reproducible and extensible framework as a foundation for future community-wide intercomparison and best-practice guidelines within CORDEX. This first phase targets two fundamental determinants of reliability under a changing climate: extrapolation (particularly for ESD) and transferability (particularly for emulators). Figure~\ref{fig:benchmark} summarizes the benchmark: (a) the three training regions; (b) training protocols for the two experiments: pseudo-reality ESD and RCM emulators, (c) inference matrix of downscaled results used for transferability, (d) general scheme of deterministic and generative downscaling setups, with predictors and predictands used in the benchmark; and (e) a schematic of generative approaches (flow-matching in this examples). 

The first CORDEX-ML-Bench experiment targets daily temperature and precipitation at $\sim 0.11^\circ$ ($\sim 10\,\text{km}$), consistent with current CORDEX protocols (Figure~\ref{fig:benchmark}a). It covers three climatically distinct regions, the European Alps, South Africa, and New Zealand, chosen for their contrasting circulation drivers, complex topography, and hydroclimatic regimes \citep{maraun2015,gutierrez2019,benchmark2026}: mid-latitude Alpine (Alps), maritime storm-track (New Zealand), and subtropical convective summer rainfall (South Africa). The framework is intentionally expandable to further domains, including tropical, polar, and monsoon regions, towards a globally comprehensive, multi-domain evaluation suite. 

The two experiments in CORDEX-ML-Bench test current challenges in ML-based downscaling through different configuraions (Figure~\ref{fig:benchmark}b,c): (i) ESD pseudo-reality experiment tests whether models trained on the historical period can extrapolate to warmer climates, and (ii) RCM emulator experiment tests interpolation and transferability for models trained in historical and future periods. Each is evaluated over several periods using a range of core metrics, summarized in \citet{rampal2026cordex}.
\begin{itemize}
  \item Cross-validation: Same GCM and setting as training (coarsened RCM predictors) but an independent historical period, testing performance under training-like conditions.
  \item Interpolation/extrapolation Same GCM, different future period. This is interpolation for the emulator, which is trained on another future period, and extrapolation for ESD.
  \item Imperfect inference: Same GCM, but with raw GCM fields instead of coarsened RCM fields as predictors.
  \item Transferability testing: Inference using a different GCM from the one used in training.
  \item Change signal evaluation: Assessment of the climate change response in future periods.
\end{itemize}
In addition, most configurations are repeated with and without orography as a predictor, to assess how much static topographic information contributes to downscaling skill.

\section{Gaps, Constraints and Future Directions}
\label{sec:gaps}

Although existing daily ML downscaling approaches already support a wide range of climate impact applications, several important research challenges remain. The CORDEX Task Force on Machine Learning has identified these as key priorities, some of which are central objectives of the Task Team. These challenges are briefly discussed below.

\subsection{Sub-daily temporal resolution}
\label{sec:subdaily}

Current emulators predominantly operate at daily timescales. Consequently, they cannot resolve sub-daily variability critical for capturing intense convective storms and flash floods \citep{lenderink2017}, diurnal wind variability and urban heat island extremes \citep{bushenkova2024,johannsen2024}. Sub-daily precipitation extremes can intensify more rapidly than daily extremes under warming, sometimes exhibiting super-Clausius--Clapeyron scaling \citep{ogorman2015,pfahl2017}. Although hourly ML downscaling methods are emerging \citep{mardani2025,pathak2024, sha2026}, their application to large ensembles remains challenging due to increased data volumes, GPU memory requirements, and limited availability of sub-daily boundary conditions from CMIP models. 
Disaggregating daily fields into hourly predictions offers a promising alternative \citep{glawion2025}, although maintaining temporal coherence across multi-day weather events remains challenging. To our knowledge, only one study has demonstrated that this approach can reproduce climate change signals in hourly extremes \citep{addison2026spatio}, which highlights an important direction for future research.

\subsection{Multivariate dependencies and compound events}
\label{sec:multivariate}

Most downscaling studies target individual prediction variables separately. Such models may not preserve inter-variable correlations in their outputs (e.g., temperature--precipitation) that underpin compound extremes, nor provide physically consistent inputs to downstream impact models (e.g., drought or flood hazard). Several multivariate emulators show promise \citep[e.g.,][]{fotiadis2024,mardani2025,schillinger2025,brochet2023multivariate}, but few studies have comprehensively tested whether they reliably reproduce inter-variable dependencies, and no consistent set of evaluation metrics exists. Inter-variable consistency is likely to matter more at finer spatial and temporal scales, as averaging to daily or coarser resolution weakens these relationships. Training multivariate approaches is challenging, particularly as the number of variables grows, because loss functions must balance variables with different error characteristics. Precipitation, for example, is typically harder to predict than temperature, so if not carefully treated models may favour some variables over others. Normalisation and error weighting may help but require further assessment.

\subsection{Convection-permitting scales}
\label{sec:cpm}

Recently, a number of generative downscaling models have been trained on convection-permitting simulations (with $\sim 1\text{--}4\,\text{km}$ grid-spacing) \citep{miralles2022,addison2022,mardani2025,fotiadis2024,addison2026}, but few have evaluated whether they reliably capture climate responses and extrapolate to warmer climates. \citet{addison2026} found that a diffusion model trained on 2.2\,km UK regional precipitation simulations (coarsened to 8.8\,km daily resolution) generated realistic fine-scale structures and intensities, and captured the main features of climate change signal in UK daily precipitation, including intensity changes up to the 99.9th percentile, albeit with some error in their magnitude. Similar errors have been reported in other studies, with emulators in some cases underestimating the climate change signal in extremes \citep{rampal2025intercomparison, rampal2024extrapolation, doury2024suitability, kendon2025}. Reducing these errors therefore remains an important challenge for emulators at convection-permitting scales.

Another key question for operational AI downscaling is whether to downscale directly from the GCM scale to the kilometre scale in a single step, or to use a nested approach, as is common in limited-area RCMs. Nested approaches appear promising, and many studies have demonstrated that it is possible to downscale from the GCM scale to an intermediate resolution \citep[e.g., 150--10\,km][]{doury2024suitability, bano2024, rampal2026downscaling}, and from intermediate to kilometre resolutions \citep[e.g., 25--2\,km][]{miralles2022, mardani2025, tomasi2025can}, but few studies have applied both steps in tandem. Other hybrid approaches combine dynamical downscaling to a coarser intermediate resolution and AI downscaling to refine from this resolution to the target kilometre scale (e.g. using super-resolution approaches).



\subsection{Urban scales}
\label{sec:cities}

Urban environments and land-use change can substantially influence local climate and weather extremes, for example through the urban heat-island effect \citep{chapman2017impact,masson2020urban}. However, the intra-urban heterogeneity in land use, building morphology and vegetation that controls these effects occurs at scales (typically below $\sim$100\,m), smaller than even convection-permitting grids. ML-based urban downscaling is emerging, with CNN and gradient-boosting models downscaling urban temperatures and heat-island intensity \citep{johannsen2024,bushenkova2024, blunn2024machine}, street-level heat-stress projections \citep{briegel2025deep}. These models can use satellite-derived descriptors, such as Local Climate Zones \citep{demuzere2022global}, building heights \citep{zhu2025globalbuildingatlas} and land surface temperature (LST) \citep{gamet2026trishna}, as predictors or targets. Because surface properties are explicit inputs, trained models could also be rerun with modified surfaces (e.g., increased tree cover or higher-albedo roofs) to explore adaptation scenarios at low cost.

However, few urban ML downscaling methods have been applied to long-term climate projections.  Beyond data limitations, ML models trained on one city may fail to transfer to others, respond unrealistically to unseen surface configurations, or, as in other ML downscaling, extrapolate poorly to warmer climates. Benchmarks assessing these problems and capturing the strongly sub-daily and multivariate urban signal (Sections~\ref{sec:subdaily}--\ref{sec:multivariate}) would further support this evaluation.

\subsection{Temporal memory and autoregressive emulators}
\label{sec:memory}

Most ML downscaling approaches predict each timestep independently and retain no memory of prior conditions unless lagged predictors or recurrent layers (e.g., LSTMs) are explicitly included. The absence of prognostic memory may impair prediction of extremes where antecedent surface conditions and land--atmosphere feedbacks are important, such as persistent heatwaves and droughts amplified by soil-moisture desiccation or snowpack anomalies \citep{kendon2025,schumacher2022,cho2025}. Training a model autoregressively is the simplest way to introduce memory, but more sophisticated approaches such as those now implemented in some neural weather models with two-way land--atmosphere coupling \citep{cho2025}, or coupled ocean--atmosphere formulations at seasonal-to-decadal timescales \citep{wang2024}, could improve such predictions further.

Following advances in global autoregressive emulators (e.g., neural weather and climate models) \citep{lang2024,perkins2025hiro,pathak2022,kochkov2024neural}, autoregressive RCM emulators are beginning to emerge, for example \citet{sha2026} trained an hourly limited-area emulator on 4\,km CONUS404 simulations driven by ERA5 boundary conditions. Unlike neural global weather and climate models, where long-term stability and multidecadal climate forcing remain major challenges, RCM emulators are constrained along their lateral boundaries by large-scale fields from the driving GCM or reanalysis, making their training and predictions more stable.


\subsection{Geographic Extrapolation}
\label{sec:geographic}
Physics-based RCMs can generally be applied to a new region simply by changing the domain, boundary conditions and model setup. In contrast, RCM emulators and ESD approaches are typically trained on data from a single region and tend to perform well only there, with skill degrading substantially when applied to regions unseen during training \citep{harder2026benchmarking,blasone2025graph}. Extrapolation across geographic domains and dynamical climate zones (such as mid-latitude storm tracks versus monsoon or tropical convective regimes) is therefore an important gap. If achievable, this would be a significant advance, facilitating the production of high-resolution climate information in data-sparse regions, including many developing regions and Pacific Island nations \citep{evans2024higher}, and would strengthen the case for ML-based downscaling as a route to more equitable participation in CORDEX.

\subsection{Transferability}
\label{sec:transferability}

As noted above, a fundamental challenge for RCM downscaling emulators is generalizing beyond the specific GCM--RCM combination and predictor configuration on which they were trained---specifically, whether a model can reliably generalize across driving GCMs with different model structures, across emission scenarios and warming regimes. Some studies have found that emulators trained on upscaled RCM predictors (``perfect'' framework) transfer better to unseen GCMs than those trained on raw GCM predictors (``imperfect'' framework), which tend to learn compensation for biases specific to the training GCM \citep{boe2023,bano2024,kendon2025}. Further testing and benchmarking would be useful, however, and all RCM emulators still inherit their target RCM's structural biases relative to observations. Promising strategies to improve cross-GCM transferability include training on multiple GCM--RCM combinations to expose the model to a broader range of boundary-condition characteristics \citep{doury2023,schillinger2025}, and developing autoregressive regional emulators that predict each regional atmospheric state from the previous state and GCM-supplied boundary conditions, allowing weather systems to evolve over time \citep{sha2026}. RCM emulators trained on historical and moderate-emissions simulations (e.g., SSP2-4.5) may struggle under warmer conditions beyond their training range (e.g., under SSP5-8.5). 

\subsection{Extreme events}
\label{sec:extremes}

Extreme weather events are of particular importance in downscaling applications because of their large societal impacts. Events with extremely rare local intensities, especially in regions of complex terrain with steep spatial variability, may be a particular challenge for ML methods. By definition, there are relatively few samples of such events to learn from, and good performance may depend on how well methods generalise beyond the training data. Climate change signals in extremes can also be difficult to learn, as they often include strong warming-driven responses \citep{GIBSON2026}. Historically, performance on extreme events with intensities associated with high impacts has not been a focus of evaluations of ML methods \citep{watson2022}. There have been an increasing number of studies that do show performance for events with intensities beyond those seen in training. Some have indicated promising performance in downscaling \citep[e.g.,][]{vosper2023,addison2026}. Studies of other ML climate science applications have also found both overall good performance \citep[e.g.,][]{watson2022,pasche2025}, and limitations of prominent ML approaches \citep[e.g.,][]{sun2024,zhang2025}. Modifications to ML methods, such as tail-weighted or extreme-value loss functions, targeted sampling of rare regimes, and generative architectures, may be necessary to achieve good performance in some situations. Process-based evaluation is also important for extremes, to test whether emulators preserve the link between local extremes and their large-scale drivers, such as circulation regimes and modes of variability (e.g., ENSO), as well as persistence-related feedbacks such as soil moisture--temperature coupling (Sections~\ref{sec:multivariate} and~\ref{sec:memory}). Overall, information about how ML performs in extreme weather situations is still too limited to form a general picture, and development of a systematic approach in benchmarks would greatly aid understanding.

\section{CORDEX Task Force on Machine Learning}
\label{sec:implementation}

\begin{figure}[tbp]
\centering
\includegraphics[width=0.9\linewidth]{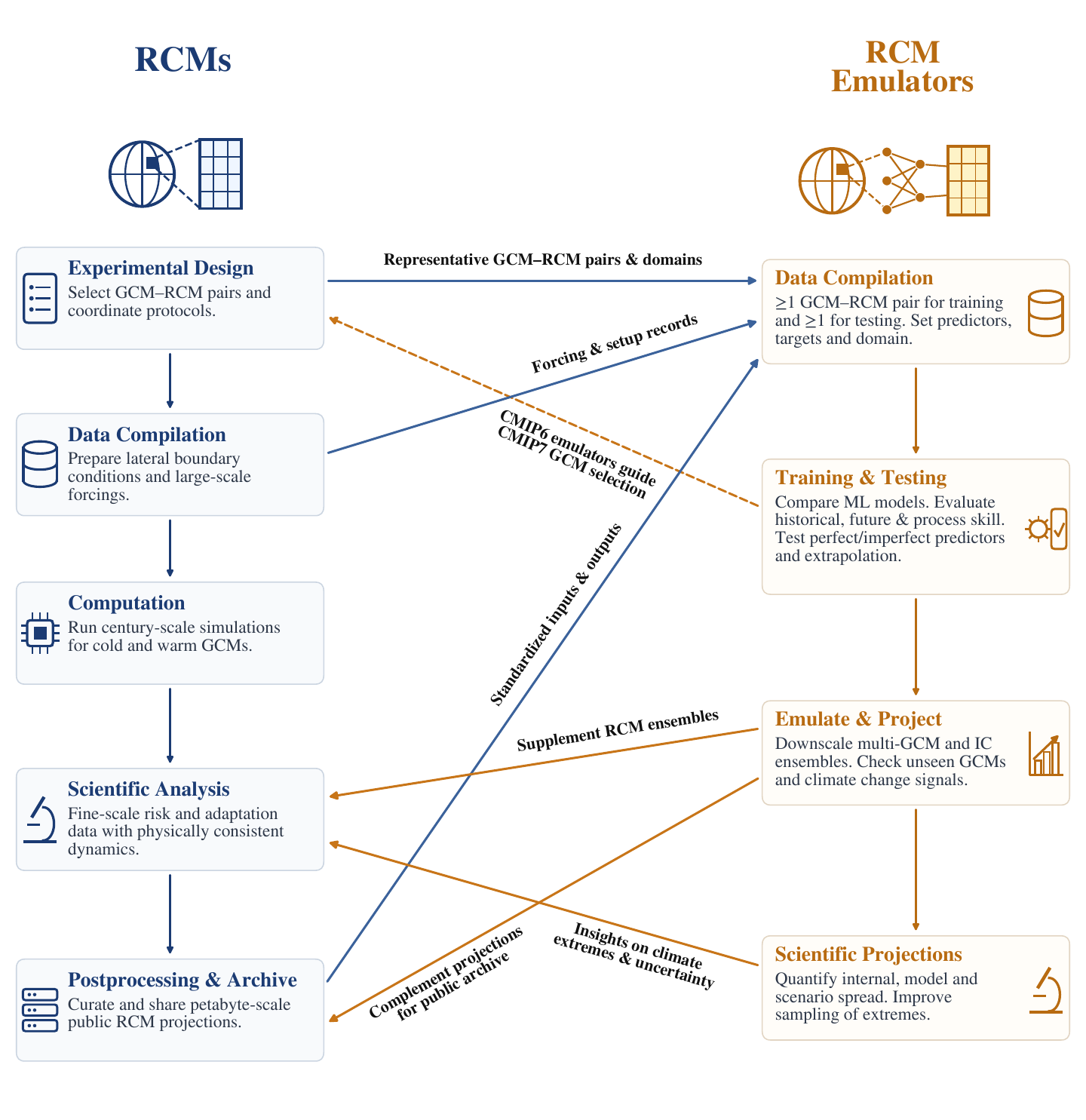}
\caption{Diagram representing the coordinated workflow of RCM downscaling emulators and conventional dynamical RCMs, adapted from \citet{vankatwyk2026}.}
\label{fig:workflow}
\end{figure}

Addressing these gaps requires a coordinated approach that leverages CORDEX's existing infrastructure while establishing new standards specific to ML-based methods. Figure~\ref{fig:workflow} illustrates a proposed hybrid framework to integrate downscaling based on ML into existing RCM workflows, spanning experimental design, data processing, and model training to projection and dissemination, outlining minimum requirements and best practices for training and evaluation. The connecting bridges in Figure~\ref{fig:workflow} show how the physics-based RCM and RCM emulator workflows interact and complement each other. RCM simulations provide the training data for emulators. In turn, emulators trained on CMIP6-driven RCM simulations can be applied across many GCMs to help select which ones to downscale dynamically for CMIP7. Large emulator ensembles can also complement existing dynamical projections by improving the sampling of climate extremes and internal variability. Sections~\ref{sec:standards}--\ref{sec:engagement} detail the specific components of this framework and introduce a proposed CORDEX Task Team on Machine Learning for Downscaling, which would continue this work by establishing standardized training best practices, defining benchmarks and reliability criteria for ML-based downscaling, and fostering coordinated community engagement across the climate and ML research communities.

\subsection{Establish Benchmarks and Standards}
\label{sec:standards}

A key activity of the TF-ML has been the development of CORDEX-ML-Bench (\autoref{sec:benchmark}), with broad participation from institutions within and beyond the Task Force. Initial results highlighted the strengths of generative models relative to deterministic approaches and identified limitations in the extrapolation capabilities of ESD approaches \citep{rampal2026cordex}. Future benchmark experiments will build on these efforts to assess model performance and address the scientific gaps identified in Section~\ref{sec:gaps}. These extensions should help establish which methods are sufficiently reliable for a range of applications, from research studies to operational climate services.

The choice of evaluation metrics is important. CORDEX-ML-Bench deliberately used a compact set of core metrics to facilitate uptake, with particular emphasis on extrapolation, following a similar approach to the first WeatherBench benchmark. Future versions should expand the core and supplementary metrics to better assess distributional fidelity, spatial dependence, rare extremes, and physical processes. Application-specific criteria should define acceptable performance, requirements for uncertainty quantification, and guidance for communicating limitations, recognizing that different uses require different levels of reliability.

\subsection{Develop Training Best Practices and Share Pretrained Models}
\label{sec:bestpractices}

Guidelines should specify minimum training data requirements, optimal predictor variable selection, and strategies for incorporating physical constraints. Particular attention should be paid to ensuring robust performance under climate change conditions and for extreme weather events, including explicit testing of extrapolation and tail performance wherever models are intended to be applied in future warming regimes beyond the training data.

Training state-of-the-art ML-based models across full continental domains and sub-daily resolutions can still require substantial GPU resources and multi-terabyte storage capacity. These requirements may challenge the computational infrastructure of institutions in several CORDEX regions, potentially reinforcing existing inequities if left unaddressed. However, rapidly falling GPU computing costs, emerging lightweight and fast-sampling architectures such as consistency models \citep{hess2025} and flow matching \citep{rampal2025intercomparison,wetherell2026flow}, and the orders-of-magnitude lower inference demands of emulator approaches compared to running dynamical RCMs can substantially enhance accessibility and help reduce these disparities in the future. Ensuring equitable access to pre-trained models, shared training datasets, and lightweight model architectures should therefore remain a strategic priority.

\subsection{Foster Community Engagement}
\label{sec:engagement}

Regular workshops and training programs should bring together climate scientists and ML researchers to share expertise and align efforts. The Task Team will address many of the gaps outlined in Section~\ref{sec:gaps} through dedicated subgroups, with broader participation from the community and other experts. This work would be supported by shared open-source software frameworks (e.g., standardized data loaders, model training templates, pre-trained models, evaluation tools, and learning materials), which would lower barriers to entry, prevent duplicated effort, and promote reproducibility across CORDEX domains. Collaboration with existing initiatives like CMIP and regional climate centers would ensure compatibility with broader climate modeling efforts.

\section{Conclusions}
\label{sec:conclusions}

The rapid advance of machine learning methods in climate science presents both an extraordinary opportunity and a significant coordination challenge for the regional climate modeling community. The CORDEX Task Force on Machine Learning was established to coordinate ML-based downscaling activities, map the evolving methodological landscape, and develop a strategic plan that balances the growing opportunities of ML-based downscaling against its remaining scientific and operational limitations.

This paper presents the Task Force's outcomes across three interconnected fronts. First, the Task Force developed CORDEX-ML-Bench, the first coordinated ML benchmarking experiment within CORDEX, bringing together 40 models from over 20 institutions across three climatically distinct regions: the European Alps, South Africa, and New Zealand \citep{rampal2026cordex}. Its initial results highlighted the strengths of generative models, particularly for precipitation, and identified important limitations in extrapolation to warmer climates with observational ESD downscaling methods. By providing publicly available datasets, standardized protocols, and clearly defined evaluation tasks, CORDEX-ML-Bench establishes a foundation for sustained community benchmarking and the reliable application of ML-based climate downscaling. Second, we proposed a common terminology to bring clarity to a field where rapid methodological growth has led to heterogeneous naming and framing. By distinguishing observation-based ESD (perfect prognosis) from RCM emulators, and static from emerging dynamic (autoregressive) and hybrid emulators, this terminology provides a coherent framework for tracking developments in the field. We also mapped the landscape of ML-based downscaling, highlighting key publications and how the field has evolved over time and across regions. Third, we identified the principal scientific gaps that currently limit reliable application, including extrapolation to warmer climates and unseen regions, transferability across driving models, representation of sub-daily extremes, multivariate dependencies, temporal memory, and systematic evaluation of rare high-impact events. These gaps define priorities for future benchmark experiments and coordinated research within CORDEX. Realizing the full potential of ML-based downscaling will require sustained coordination on benchmarking, training best practices, equitable access to tools and datasets, and closer collaboration between the climate science and ML communities. The proposed follow-up Task Team would build on these outcomes to support the long-term development and responsible application of ML-based climate downscaling.

\section*{Data and Code Availability}
The open-access CORDEX-ML-Bench dataset is archived on Zenodo at \url{https://doi.org/10.5281/zenodo.17957264} \citep{benchmarkdata2025} and is also available on Hugging Face at \url{https://huggingface.co/datasets/rampaln/CORDEX-ML-Bench}. Benchmark protocols, code, and baseline evaluation workflows are hosted on GitHub at \url{https://github.com/WCRP-CORDEX/ml-benchmark} \citep{rampal2026cordex}.

\section*{Acknowledgments}

N.R. and P.B.G. received funding from the New Zealand MBIE Endeavour Smart Ideas Fund (NIW2504). J.M.G. and J.B.-M. acknowledge support from the Copernicus Climate Change Service (C3S), implemented by ECMWF on behalf of the European Union under contract C3S2\_384. M.L.B. acknowledges support from CREATOR-ANR-25-CE56-3663. R.F.-F. acknowledges support from the Bolin Centre for Climate Research and the Horizon Europe AI4PEX project (grant 101137682). A.O. acknowledges support from UKRI/NERC, including the grant Drivers and Impacts of Extreme Weather Events in Antarctica (ExtAnt; NE/Y503307/1) and the NERC National Capability International grant SURface FluxEs In AnTarctica (SURFEIT; NE/X009319/1). P.M.M.S. acknowledges funding from FCT, I.P./MCTES through national funds (PIDDAC): LA/P/0068/2020 and UID/50019/2025, and from the European Union NextGenerationEU projects UID/PRR/50019/2025 and UID/PRR2/50019/2025. P.A.G.W. was supported by Natural Environment Research Council grant NE/Z000076/1.

\bibliographystyle{plainnat}
\bibliography{references}

\nolinenumbers

%
%


\end{document}